\documentclass[12pt]{iopart}

\usepackage{iopams}
\usepackage{graphicx}
\begin{document}

\title[A systematic study of the factors affecting central depletion in nuclei]{A systematic study of the factors affecting central depletion in nuclei}

\author{G. Saxena$^{1,2,\S}$, M. Kumawat$^{1,2}$, B. K. Agrawal$^{3,4,\dagger}$ and Mamta Aggarwal$^{5}$}

\address{$^{1}$Department of Physics, Govt. Women Engineering College, Ajmer-305002, India}
\address{$^{2}$Department of Physics, Manipal University Jaipur, Jaipur-303007, India}
\address{$^{3}$Saha Institute of Nuclear Physics, 1/AF Bidhannagar, Kolkata-700064, India}
\address{$^{4}$Homi Bhabha National Institute, Anushakti Nagar, Mumbai-400094, India}
\address{$^{5}$Department of Physics, University of Mumbai, Kalina, Mumbai-400098, India.}
\ead{$^{\dagger}$bijay.agrawal@saha.ac.in, $^{\S}$gauravphy@gmail.com}
\vspace{10pt}
\begin{indented}
\item[]Jan 2019
\end{indented}

\begin{abstract}
A systematic study of the central depletion of proton density has been
performed in the isotonic chains of nuclei with neutron numbers $N = 20$
and $28$ using different variants of the relativistic mean-field (RMF)
models. These models include either the non-linear contributions from
the mesons with the coupling constants being density independent or
the non-linearity of the mesonic fields realized through the density
dependent coupling strengths. The central depletion in deformed nuclei tends to disappear irrespective of the occupancy of
$2s_{1/2}$ state in contrast to the spherical nuclei in which the unoccupancy of $2s_{1/2}$ state leads to
the central depletion. Due to the differences in the strength of spin-orbit potentials in these models, the central depletions are found to be model dependent. The influence of the central depletion on the neutron-skin thickness is also investigated. It appears that the effects of the central depletion do not percolate far
enough to display its finger prints on the trends of the neutron-skin thickness.\end{abstract}

\submitto{\JPG}

\section{Introduction}

The "Bubble Structure" or the depletion in the central density of nucleons has attracted a lot of research interest currently especially due to the availability of the advanced experimental facilities to study the exotic nuclei. The increasing experimental \cite{Mutschler16} and theoretical efforts \cite{Todd04,Grasso07,Duguet16,Schuetrumpf17,Li16,Phuc2018,Saxena17,Saxena18} in order to search and understand the bubble like structures have provided a significant amount of information on the bubble nuclei. This phenomenon of bubble is usually believed to be due to the unoccupancy of the $s_{1/2}$-state which leads to significant reduction of the density at the center. Sometimes, the bubble or the depletion in the central density is associated with the inversion of ($2s_{1/2}$ and 1d$_{3/2}$) states and ($3s_{1/2}$ and $1h_{11/2}$) states ~\cite{Khan07} causing unoccupancy of $s_{1/2}$-state. In case of heavy and superheavy nuclei \cite{Saxena18,Sobiczewski2007,Decharge99,Singh12,Ikram14}, the appearance of the bubble structures has been attributed to the effects of electrostatic repulsion and the symmetry energy ~\cite{Schuetrumpf17}.\par

Experimental signature of empty $2s_{1/2}$ in $^{34}$Si using one-proton removal reaction technique, recently reported by Mutschler \textit{et
al.}~\cite{Mutschler16}, has opened a testing ground for the already
developed successful models and new avenues for the nuclear structure
mechanisms related to the nucleonic central density depletion across
the periodic chart. On the theoretical side, various models like the
nuclear density functional theory \cite{Schuetrumpf17}, ab-initio
self consistent Green's function many-body method \cite{Duguet16},
relativistic mean-field and non-relativistic mean-field
models along with shell-model \cite{Grasso09}, and the particle-number and angular-momentum
projected Generator Coordinate Method based on Hartree-Fock-Bogoliubov
+ Lipkin-Nogami states with axial quadrupole deformation \cite{Yao12}
etc., have been applied to study the bubble like structures. The nuclei
$^{34}$Si and $^{22}$O are suggested to be the strong candidates for the
proton and neutron bubbles, respectively. The tensor force effect and the
pairing correlations in the bubble structure have been also investigated
~\cite{Khan07,Wu13,Wang11,Nakada12,Wang11a}. A recent work has reported
the central depletion in the deformed sd-shell nuclei \cite{Shukla2014}
where the extent of central depletion is generally found to be weaker than
those in the spherical nuclei in similar mass region. The weakening of the
central depletion in deformed nuclei may be due to the change in the occupancy
of $s_{1/2}$-state. The actual mechanism of the weakening of central depletion in
deformed nuclei has not been investigated in detail so far.
\par

Henceforth, a study towards the deeper understanding and clarity on the
various issues regarding the mechanisms behind the bubble phenomena
discussed above, which is the need of the current research interest,
is the objective of the present work. Here we perform a systematic study
of the central depletion in the nuclei with neutron numbers $N = 20$ and
$28$ using relativistic mean-field (RMF) models. The RMF models employed in
this work are: (a) models including  the contributions from the nonlinear
self- and mixed-interactions of the mesons with various coupling strengths
taken to be constant \cite{walecka74,Boguta77,Boguta83,Furnstahl97} and (b) the
models with only linear contributions from the mesons, the coupling
strengths are density-dependent \cite{Todd-Rutel05}. The well deserved
attention has been paid to the effects of deformation for which we
study well deformed nuclei $^{40}$Mg, $^{42}$Si and $^{44}$S that have
shown central density depletion. Also, we perform a systematic study
of $^{24-48}$Si isotopes that are found with oblate, spherical as well
as prolate shapes with a range of ($\beta = 0-0.4$) deformation, hence
provide an ideal testing ground to study central depletion variation
with changing deformations and shapes. We also examine the influence of
neutron to proton ratio and pairing energy in addition to occupancy in
$2s_{1/2}$-state on the bubble effect. The influence of central depletion
on the variation of neutron-skin thickness with the asymmetry has been
explored.  \par

In section II, we briefly describe the different RMF models
considered. The main results are presented in section III. In section IV,
we present our conclusions.

\section{Theoretical Formalisms} The effective lagrangian for the RMF models can be broadly classified in two categories: (i) in which linear terms for the mesons coupled to the nucleonic degrees of freedom and non-linear terms for mesons describe their self and mixed interactions, the coupling constants independent of the density \cite{walecka74,Boguta77,Boguta83}, and (ii) contains only linear terms for the mesons and the non-linear contributions of mesons are accounted through the density dependence of the coupling constants. We use the RMF models belonging to both the categories. We consider the NL3 and NL3* parameterizations \cite{Lalazissis97,Lalazissis09} of the RMF model which include linear terms for the $\sigma$, $\omega$ and $\rho$ mesons and non-linear term only for the self interaction of the $\sigma$ meson. The FSU-Gold and FUS-Garnet parameterizations include in addition the non-linear self interaction of $\omega$ meson and the mixed interaction terms for $\omega$ and $\rho$ mesons \cite{Todd-Rutel05,Chen15}. The RMF model belonging to the category of density dependence of coupling constants for the meson exchange are the DD-ME2 and DD-PC1 \cite{Lalazissis05,Niksic08}. The effective lagrangian for DD-PC1 model is analogous to DD-ME2 model but, it does not include the derivative term for mesonic fields  and hence they are directly expressed in terms of nucleonic field.\par

The interaction part of the effective Lagrangian of RMF model belonging to the category
(i) can be written as,

\begin{eqnarray} \label{eq:Lagrangian2}
{\cal L}_{\it int}=&&\overline{\psi}\left [g_{\sigma} \sigma -\gamma^{\mu} \left (g_{\omega }
\omega_{\mu}+\frac{1}{2}g_{\mathbf{\rho}}\tau .
\mathbf{\rho}_{\mu}+\frac{e}{2}(1+\tau_3)A_{\mu}\right ) \right ]\psi
-\frac{{\kappa_3}}{6M}
g_{\sigma}m_{\sigma}^2\sigma^3\nonumber \\&&-\frac{{\kappa_4}}{24M^2}g_{\sigma}^2
m_{\sigma}^2\sigma^4+ \frac{1}{24}\zeta_0
g_{\omega}^{2}(\omega_{\mu}\omega^{\mu})^{2}
+\frac{\eta_{2\rho}}{4M^2}g_{\omega}^2m_{\rho
}^{2}\omega_{\mu}\omega^{\mu}\rho_{\nu}\rho^{\nu}
\end{eqnarray}
Where the symbols have usual meaning and the details can be found in Refs.$~$ \cite{Boguta77,Boguta83,Furnstahl97,Todd-Rutel05}.

The density-dependent meson-exchange model (DD-ME) \cite{Lalazissis05} interaction part of the Lagrangian
does not contain any non-linear term, but, the meson-nucleon strengths
$g_{\sigma}$, $g_{\omega}$ and $g_{\rho}$ have an explicit density
dependence in the following form:

\begin{equation} \label{eq:Equation1} 
g_{i}(\rho) =  g_{i}(\rho_{sat})f_{i}(x), \,\,\,\,\,\,\, for \,\,\, i = \sigma, \omega 
\end{equation}

where the density dependence is given by
\begin{equation} \label{eq:Equation2} 
f_{i}(x) =  a_{i} \frac {1+b_{i}(x+d_{i})^{2}}{1+c_{i}(x+e_{i})^{2}} 
\end{equation}
in which $x$ is given by $x = \rho/\rho_{sat}$, and $\rho_{sat}$ denotes the baryon density
at saturation in symmetric nuclear matter.
For the $\rho$ meson, density dependence is of exponential form and given by
\begin{equation} \label{eq:Equation3} 
f_{\rho}(x) =  exp(-a_{\rho}(x-1)) 
\end{equation}

Interaction part of Lagrangian for the density-dependent point coupling model (DD-PC) \cite{Niksic08} is given by
\begin{eqnarray} \label{eq:Lagrangian3} 
{\cal L}_{\it int}=&& -\frac{1}{2}\alpha_{S}(\rho)(\overline{\psi}\psi)(\overline{\psi}\psi) -
\frac{1}{2}\alpha_{V}(\rho)(\overline{\psi}\gamma^{\mu}\psi)(\overline{\psi}\gamma_{\mu}\psi)\nonumber
\\ &&- \frac{1}{2}\alpha_{TV}(\rho)(\overline{\psi}\overrightarrow{\tau}\gamma^{\mu}\psi)(\overline{\psi}\overrightarrow{\tau}\gamma_{\mu}\psi) - \frac{1}{2}\delta_{S}(\rho)(\overline{\psi}\psi)\square(\overline{\psi}\psi) \nonumber 
\end{eqnarray}

In analogy with meson-exchange model (DD-ME) described above, this model
contains isoscalar-scalar (S), isoscalar-vector (V) and isovector-vector
(TV) interactions. The coupling constants $\alpha_{i}(\rho)$ are density
dependent and have the form \cite{Niksic08}:

\begin{equation} \label{eq:Equation4} 
\alpha_{i}(\rho)  =  a_{i} + (b_{i} + C_{i}x)e^{-d_{i}x}, for \,\,\, i = S, V, TV
\end{equation}

Some characteristic parameters associated with
nuclear matter at the saturation density are listed in Table \ref{tab1}
for the RMF models considered. Using these RMF models we aim to examine
the influence of various factors on the central depletion. \par

\begin{table*}
\caption{Nuclear matter properties as saturation density $\rho_{sat}$, binding energy per nucleon $\epsilon$, effective nucleon mass $m^{*}/m$, incompressibility coefficient $K$, symmetry energy $J$ and slope of symmetry energy $L$ at the saturation density are given for various RMF models.}
\centering
\resizebox{0.8\textwidth}{!}{%
\begin{tabular}{c|c|c|c|c|c|c}
\hline
\multicolumn{1}{c|}{Nuclear Matter Properties}&\multicolumn{1}{|c|}{NL3}&\multicolumn{1}{|c|}{NL3$^*$}&\multicolumn{1}{|c|}{FSU-Gold}&\multicolumn{1}{|c|}{FSU-Garnet}&\multicolumn{1}{|c|}{DD-ME2}&\multicolumn{1}{|c}{DD-PC1}\\
\hline
$\rho_{sat}$(fm$^{-3}$)  &0.149         &0.150   &0.148  &0.153     &0.152   &0.152       \\
$\epsilon$(MeV) &-16.30        &-16.31 &-16.30  &-16.23    &-16.14  &-16.06      \\
$m^{*}/m$    &0.60            &0.59   &0.61   &0.58     &0.57    &0.66        \\
$K$(MeV)           &271.76         &258.78    &230.00    &229.50     &250.89  &230.00         \\
$J$(MeV)               &37.4           &38.6  &32.6  &30.9     &32.3    &33.0          \\
$L$(MeV)               &118.6         &122.6  &60.5   &51.0     &51.2    &57.2        \\
\hline
\end{tabular}}
\label{tab1}
\end{table*}

The degree of depletion in the proton or neutron densities at the center of the
nuclei is usually expressed in terms of the so-called bubble parameters. These
parameters are not uniquely defined. We have employed a simple definition of the bubble parameter \cite{Grasso07},
\begin{equation}
b_\tau = 1-\frac{\rho_{\tau,c}}{\rho_{\tau,max}}
\label{bt}
\end{equation}
where, $\tau= p,n$ and $\rho_{\tau,c}$, $\rho_{\tau,max}$ represent the central density, maximum density respectively. We also adopt the bubble parameter $b^{\prime}_\tau$
defined as \cite{Schuetrumpf17},
\begin{equation}
b^{\prime}_{\tau} =1-\frac{\rho_{\tau,c}}{\rho_{\tau,av}}
\label{btp}
\end{equation}
where, $\rho_{\tau,av}$ is the average density defined as
\begin{equation}
\rho_{\tau,av}=\frac{3 N_\tau}{4\pi R_{\tau,d}^3}
\end{equation}
with $R_{\tau,d}$ being the diffraction radius \cite{Schuetrumpf17,Friedrich82}
and $N_\tau$ is nucleon number. It is evident from Eqs. in (\ref{bt} and \ref{btp}) that the parameters $b_\tau$ and $b^\prime_\tau$ estimate the extent of depletion in the central density with respect to the maximum and the average
densities, respectively. Thus, their values are not expected to be identical for
 a given nucleus and the model. The positive values of $b_\tau$ and
$b^{\prime}_\tau$ indicate the existence of the central depletion.\par

\section{Results and discussions}
We now present our results for the systematics of central depletion
in the isotonic chains of the nuclei with neutron numbers $N = 20$ and
$28$. The results are obtained using different variants of the RMF model
as outlined briefly in the previous section. The models employed are
namely  NL3, NL3$^*$, FSU-Gold, FSU-Garnet, DD-ME2 and DD-PC1 (see Table
\ref{tab1}). We will examine here influence of various factors viz. model
dependency, occupation of $2s_{1/2}$, neutron to proton ratio, deformation
and pairing contribution, on the central depletion. A systematic study of
Si isotopes ($N = 10-34$) exhibiting a range of deformations and shapes
is presented. We also look for the imprints of the central depletion
on the systematics of the neutron-skin thickness. In particular, the
effects of central depletion on the variation of neutron-skin thickness
with asymmetry parameter $(N-Z)/A$ is investigated.

\begin{table*}
\caption{A comparison of the values of bubble parameters $b_p$ and
$b^{\prime}_p$ obtained using different RMF models.}
\centering
\resizebox{1.0\textwidth}{!}{%
\begin{tabular}{c|c|c|c|c|c|c|c|c|c|c|c|c|c}
\hline
\multicolumn{1}{c|}{Isotones}&\multicolumn{1}{|c|}{Bubble}& \multicolumn{6}{|c|}{$b_p$}&\multicolumn{6}{|c}{$b^{\prime}_p$}\\
\cline{3-14}
\multicolumn{1}{c|}{with N}&\multicolumn{1}{|c|}{Nucleus}& \multicolumn{1}{|c|}{NL3}&\multicolumn{1}{|c|}{NL3$^*$}&\multicolumn{1}{|c|}{FSU-Gold}&\multicolumn{1}{|c|}{FSU-Garnet}&\multicolumn{1}{|c|}{DD-ME2}&\multicolumn{1}{|c|}{DD-PC1}
&\multicolumn{1}{|c|}{NL3}&\multicolumn{1}{|c|}{NL3$^*$}&\multicolumn{1}{|c|}{FSU-Gold}&\multicolumn{1}{|c|}{FSU-Garnet}&\multicolumn{1}{|c|}{DD-ME2}&\multicolumn{1}{|c}{DD-PC1}\\
\hline
20&$^{30}$Ne   &0.21&0.19&0.27&0.23&0.23&0.17                            &0.07&0.05&0.18&0.16 &0.17 & 0.00\\
               &$^{32}$Mg   &0.25&0.24&0.32&0.28&0.29&0.20                            &0.17&0.15&0.27&0.25 &0.41 & 0.37\\
               &$^{34}$Si   &0.35&0.33&0.41&0.38&0.37&0.27                            &0.29&0.28&0.37&0.35 &0.32 & 0.23\\\hline
28&$^{46}$Ar   &0.34&0.28&0.45&0.39&0.31&0.09                            &0.20&0.15&0.32&0.29 &0.20 &-0.02\\\hline
40&$^{56}$S    &0.30&0.60&0.45&0.35&0.23&0.00                            &0.09&0.51&0.30&0.22 &0.07 &-0.13\\
               &$^{58}$Ar   &0.47&0.62&0.35&0.20&0.00&0.00                            &0.34&0.54&0.20&0.08 &-0.07&-0.23\\
\hline
\end{tabular}}
\label{tab2}
\end{table*}

\begin{figure}[h]
\centering
\includegraphics[width=0.6\textwidth]{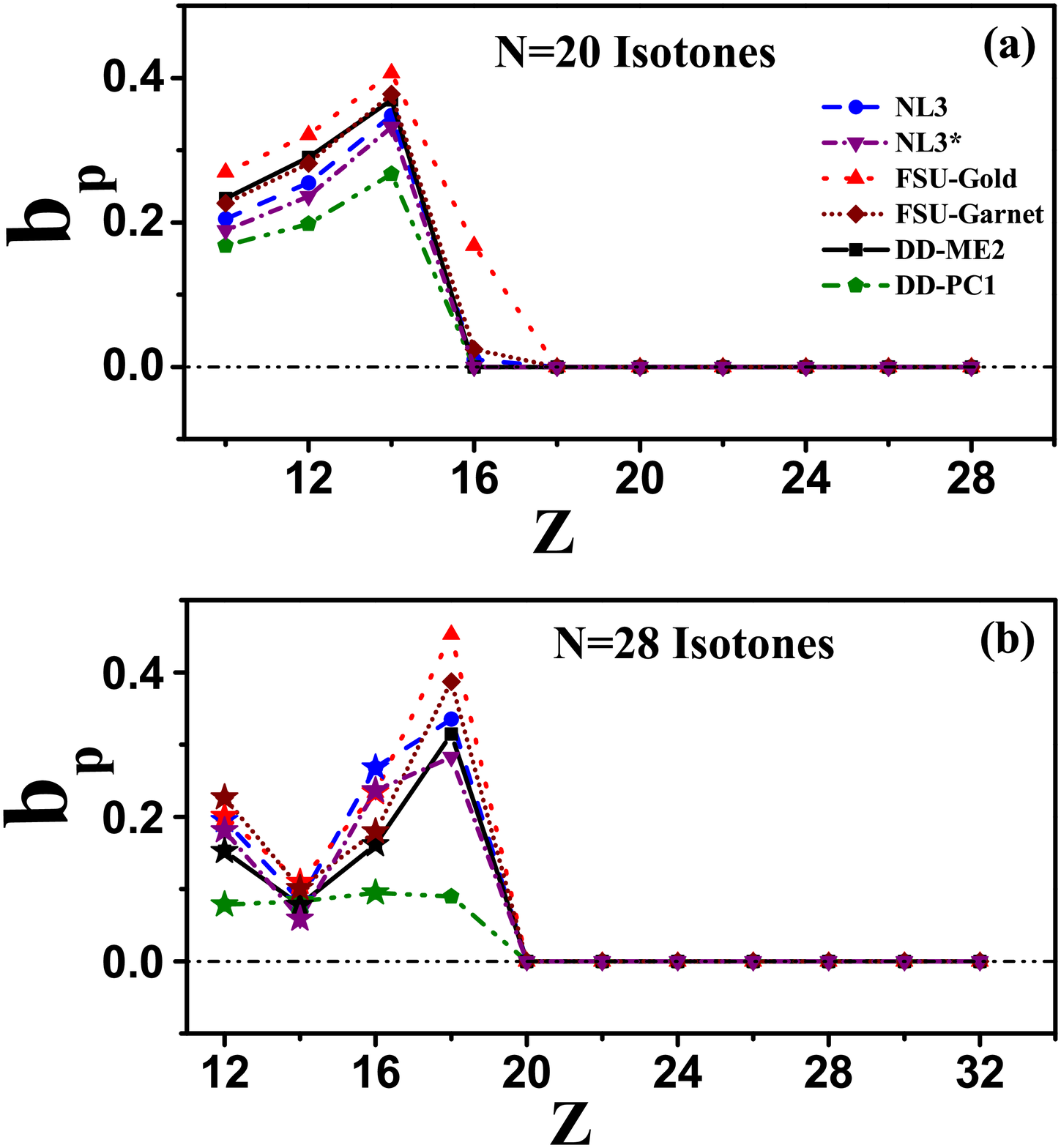}
\caption{(Colour online) The bubble parameter $b_p$ for protons in isotonic chain with neutron numbers $N = 20$ and $28$ obtained by RMF (NL3, NL3$^*$,FSU-Gold, FSU-Garnet, DD-ME2 and DD-PC1 force parameters). The asterisk symbols highlight the deformed nuclei in the case of $N = 28$ isotones.}
\label{fig1}
\end{figure}

We compare in Table \ref{tab2} the results for the $b_p$ and
$b^{\prime}_p$ (Eqs. \ref{bt} and \ref{btp}) for some selected spherical
nuclei with $N = 20, 28$ and $40$ obtained for several RMF models. One
can see that the values of $b_p$ are larger by 20\%-30\% as compared
to the $b^{\prime}_p$, because the maximum density $\rho_{\tau,max}$
is always larger than average density $\rho_{\tau,av}$. Bubble parameter
$b_p$ is easy to calculate for the spherical as well as for the deformed
nuclei. Therefore, here after, we restrict ourselves to the calculations
of bubble parameter $b_p$ only. Fig. \ref{fig1} shows our calculated
bubble parameters $b_p$ for $N = 20 $ and $28$ isotonic chains obtained
for different RMF models considered in this work. All the nuclei belonging
to $N = 20$ isotonic chain are found to be spherical for all the RMF
models employed. For $N = 28$ isotonic chain, the nuclei $^{40}$Mg,
$^{42}$Si and $^{44}$S are found to be deformed in agreement with the
available experimental predictions \cite{nndc}.

\begin{figure}
\centering
\includegraphics[width=0.6\textwidth]{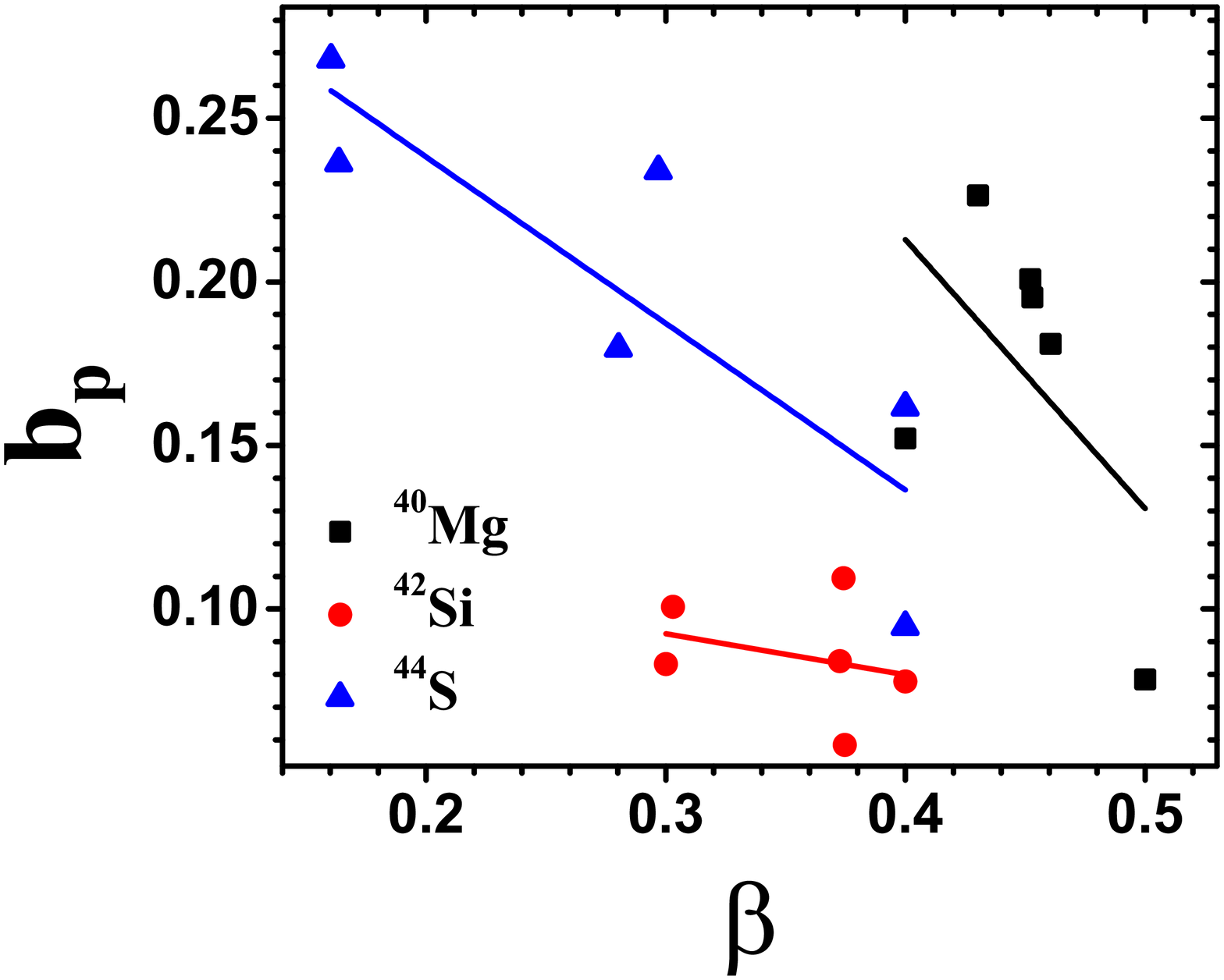}
\caption{(Colour online) The bubble parameter $b_p$ as a function of deformation $\beta$ obtained by various models for $^{40}$Mg, $^{42}$Si and $^{44}$S.}
\label{fig2}
\end{figure}

It is evident from the figure that the degree of central depletion is
sensitive to the choice of model. For the case of $N = 20$ isotones, the
central depletion of proton density is seen in nuclei $^{30}$Ne, $^{32}$Mg
and $^{34}$Si. For these nuclei, the values of the bubble parameter
for most of the models are in the range of $\sim 0.2 - 0.4$ whereas the
values of $b_p$ for the DD-PC1 model are relatively smaller. For $N = 28$
isotones, except for the DD-PC1 model, the significant central depletion
is observed in the proton density of spherical nucleus $^{46}$Ar, central
depletion seen in $^{34}$Si and $^{46}$Ar in the present work agrees with
that indicated by experimental ~\cite{Mutschler16} and earlier theoretical
~\cite{Duguet16,Schuetrumpf17,Li16,Phuc2018,Khan07,Grasso09,Wang11,Wang11a}
works. For the deformed nuclei $^{40}$Mg, $^{42}$Si and $^{44}$S the
central depletion is observed, however, this depletion is noticeably
smaller as compared to neighbouring spherical nuclei. The weakening of
central depletion in deformed nuclei is also observed in $^{24}$Ne,
$^{32}$Si and $^{32}$Ar as reported in Ref. \cite{Shukla2014}. The
prolately deformed nuclei $^{40}$Mg and $^{44}$S exhibit somewhat larger
central depletion in comparison to the oblate nucleus $^{42}$Si. The
spread in the values of $b_p$ for in $^{40}$Mg, $^{42}$Si and $^{44}$S
may be partly due to the variation in the deformation parameter obtained
for different models.

Fig. \ref{fig2} shows variation of $b_p$ with
respect to quadrupole deformation parameter $\beta$. Straight lines are
plotted to guide the eye. By and large, one can conclude that the bubble
parameter decreases with increasing deformation. In the case of prolate
deformation ($^{40}$Mg and $^{44}$S) the $b_p$ decreases rapidly with
$\beta$ as compared to oblate nucleus ($^{42}$Si). It appears that the
deformation tends to quench the central depletion. More on this will be
discussed below.\par

\begin{figure}
\centering
\includegraphics[width=0.6\textwidth]{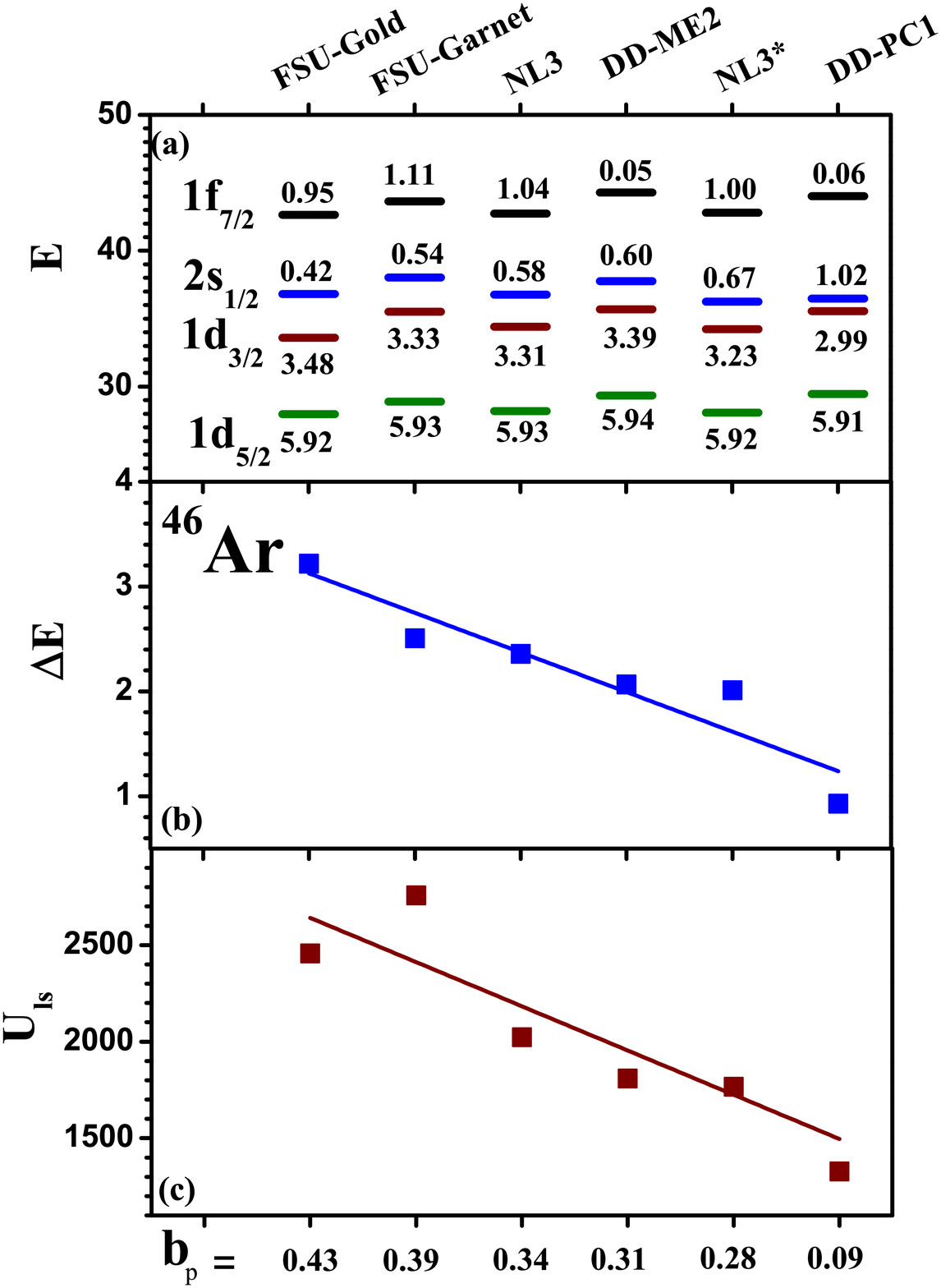}
\caption{(Colour online) (a) Proton s.p. energies of $^{46}$Ar, with respect to energy of $1s_{1/2}$ state, along with their occupancies from several models considered. Values of $b_p$ by these models are also indicated. (b) energy gap between $2s_{1/2}$ and $1d_{3/2}$ for $^{46}$Ar (c) strength of the spin-orbit potential $U_{ls}$ \cite{sharma95,satula} for various models/parameters. (The energies are in the unit of MeV).}
\label{fig3}
\end{figure}

Since the central depletion in the nuclei is believed
to be associated with the unoccupancy of $s_{1/2}$ state
\cite{Duguet16,Khan07,Grasso09,Wang11,Wang11a}, it is important
to investigate whether one can owe the differences in the extent of
central depletion across the various models to the differences in the
several inputs of the models. We consider the case of $^{46}$Ar nucleus
which shows strong model dependence \cite{Song2015} in the values of
bubble parameter. We display in Fig. \ref{fig3}, the single particle
(s.p.) energies of $^{46}$Ar along with their occupancies calculated
for several models used for the present study. The values of the bubble
parameter for a given model are indicated in the figure. It
is evidently seen that the bubble parameter $b_p$, quantifying the
central depletion, decreases as the occupancy of $2s_{1/2}$ state
increases. The occupancy of $2s_{1/2}$ state is found maximum for DD-PC1
force parameter comparative to other considered models/parameters. This
difference is examined by taking into account the various inputs of the
considered models. Among the inputs, coupling constants describing the interaction between mesons and nucleons are found to play a dominant role.
These coupling constants predominantly determine the strength of the spin-orbit potential \cite{sharma95,satula} given by
\begin{equation}
U_{ls} \varpropto \frac{1}{m^{*}/m} (C_{\sigma}^{2} + C_{\omega}^{2} + C_{\rho}^{2})
\end{equation}
where $m^{*}/m$ is the nucleon (effective) mass; the constants $C_{i}$ (i = $\sigma$, $\omega$, $\rho$) are defined as
$C_{i} = mg_{i}/m_{i}$, $m_{i}$ and $g_{i}$ being the meson masses and coupling constant, respectively. For DD-ME2 and DD-PC1 models these coupling constant are evaluated at saturation density. In Fig. \ref{fig3}(c) the spin orbit potential strength ($U_{ls}$) is plotted against bubble parameter. The different values of $U_{ls}$ for different models are responsible for the differences in the relative separation of 2s and 1d states and hence for the differences in the pairing properties.
Further, the reasonable correlation between $U_{ls}$ and bubble parameter provides plausible explanation for the existence of the model dependence.
The differences in the spin-orbit splitting may be partially attributed to the differences in the deformation and hence the central depletion, as also seen in Fig. \ref{fig2}. \par

\begin{figure}
\centering
\includegraphics[width=0.6\textwidth]{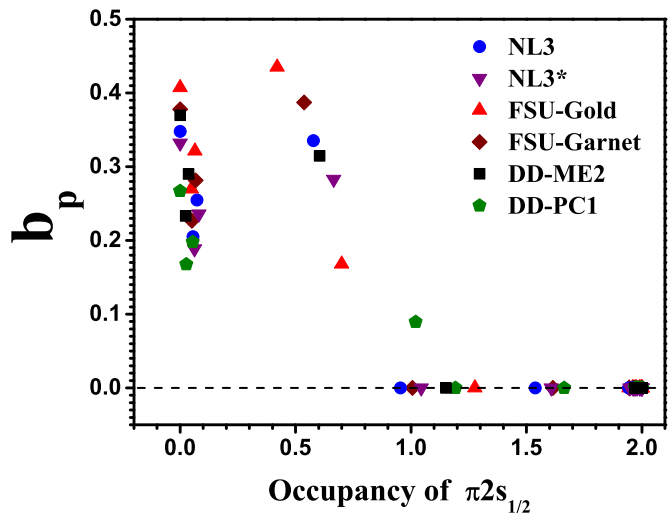}
\caption{(Colour online) Variations of bubble parameter $b_p$ as a function
of the occupancy of proton $2s_{1/2}$ state for the spherical nuclei.} \label{fig4}
\end{figure}

To visualize the dependence on occupancy of $2s_{1/2}$ state, we show
in Fig. \ref{fig4}, the variation of the bubble parameter $b_p$ plotted
against the occupancy of $2s_{1/2}$ state for protons, obtained using
different RMF models for the spherical nuclei corresponding to $N = 20$
and $28$ isotonic chains.  The overall trend suggests that the bubble
parameter decreases as the occupancy of $2s_{1/2}$ increases. With the
increase of occupancy of $2s_{1/2}$ state to 50$\%$ (i.e. $2s_{1/2}$ state
has one nucleon), $b_p$ collapses to close to zero. However, many of the
nuclei having almost unoccupied $2s_{1/2} $ state, show noticeable spread
in the values of bubble parameter. For instance, $b_p$ for $^{34}$Si
nucleus varies in the range of 0.27 to 0.41 although the $2s_{1/2}$
state is practically unoccupied for all the models. This means that
there may be other factors as well which influence the central depletion
in addition to the choice of model and occupancy of the s-orbit. Thus,
the condition that $2s_{1/2}$ state must be practically unoccupied is
only a necessary condition but not the sufficient condition for the
occurrence of the central depletion.\par

\begin{figure}
\centering
\includegraphics[width=0.6\textwidth]{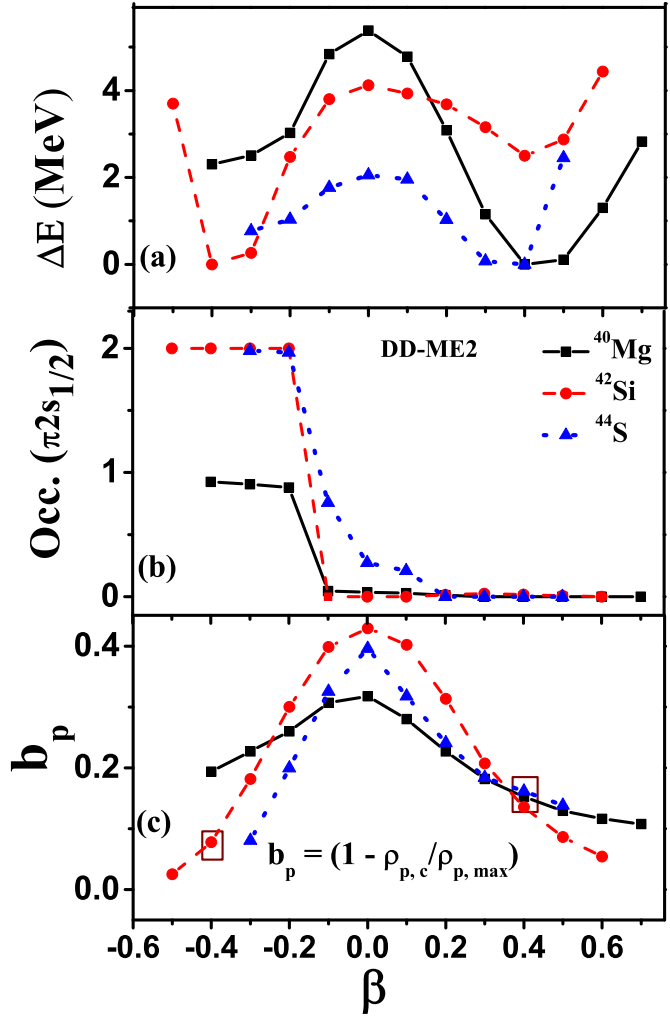}
\caption{(Colour online) (a) Binding energy (MeV) (b) occupancy of proton $2s_{1/2}$ state and (c) bubble parameter $b_p$ as a function of deformation $\beta$ for $^{40}$Mg, $^{42}$Si and $^{44}$S.}
\label{fig5} \end{figure}

Nuclear deformation is one of the factors which is
expected to play a key role in the central depression
~\cite{Duguet16,Khan07,Grasso09,Yao12,Wu13}. We investigate the quenching
of central depletion due to deformation in detail for which we consider
the cases of well deformed $^{40}$Mg, $^{42}$Si and $^{44}$S (see also
Figs. \ref{fig1} and \ref{fig2}). We plot (a) the binding energy,
(b) the occupancy of the $2s_{1/2}$ state for the protons and (c) the
bubble parameter $b_{p}$, as a function of the quadrupole deformation
parameter $\beta$ varying from -0.6 to 0.6 computed using the DD-ME2
model in Fig. \ref{fig5}. The value of deformation at each point
of the curve is obtained by constraining the quadrupole moment through
the variational procedure \cite{Niksic14}.
The trend for the other RMF models (not shown here) have been found to be qualitatively
similar to those shown in Fig. \ref{fig5}. The potential energy surface
as displayed in the top panel shows energy minima at $\beta = 0.4$ for
$^{40}$Mg and $^{44}$S, and $\beta = -0.4$ for $^{42}$Si, characterizing
prolate and oblate shapes, respectively. The bubble parameter $b_p$
shown in Fig. \ref{fig5}(c) is maximum at $\beta = 0$ and decreases as
$\beta$ increases on both prolate and oblate sides which shows quenching
of central depletion due to deformation. From Fig. \ref{fig5}(b), it can
be seen that the $2s_{1/2}$ state is highly occupied towards oblate side
and it is almost unoccupied towards prolate side. It appears that the
occupancy in s-orbit appears to be not playing a major role in variation
of $b_p$. The $b_p$ for these nuclei is decreasing on the prolate side
even though the occupancy in $2s_{1/2}$ is vanishingly small. The value
of $b_p$ (highlighted by squares in \ref{fig5}(c)) corresponding to the
respective energy minima for $^{40}$Mg, $^{44}$S and $^{42}$Si, are found
to be $b_p = 0.15$, $0.16$ and $0.08$, respectively, even the absolute
value of $\beta$ is almost same for these nuclei. The lower value of $b_p$
in $^{42}$Si (oblate) could be due to combined effect of deformation and
occupancy of $2s_{1/2}$ state. Hence, deformation seems to play the
predominant role in determining the $b_p$ as compared to the occupancy
of $2s_{1/2}$ state. This could be due to the lowering of some of the deformed single particle states.\par

Moreover, since bubble effect is related to the shell effects
associated with the unoccupancy of s-orbital surrounded by single particle
states with larger orbital angular momentum
well separated in energy  to ensure the
weak dynamical correlations. The increase in deformation
leads to stronger dynamical correlations, overlapping of sd-states
and also less pronounced shell effect \cite{moya} which consequently
disfavors formation of bubble. For better insight about the effect of
deformation on bubble structure one possible way is to calculate s-wave
projections of the nucleon wave functions and densities in the deformed
cases \cite{moya,zamick} which is left for our future work.

\begin{figure}
\centering
\includegraphics[width=0.6\textwidth]{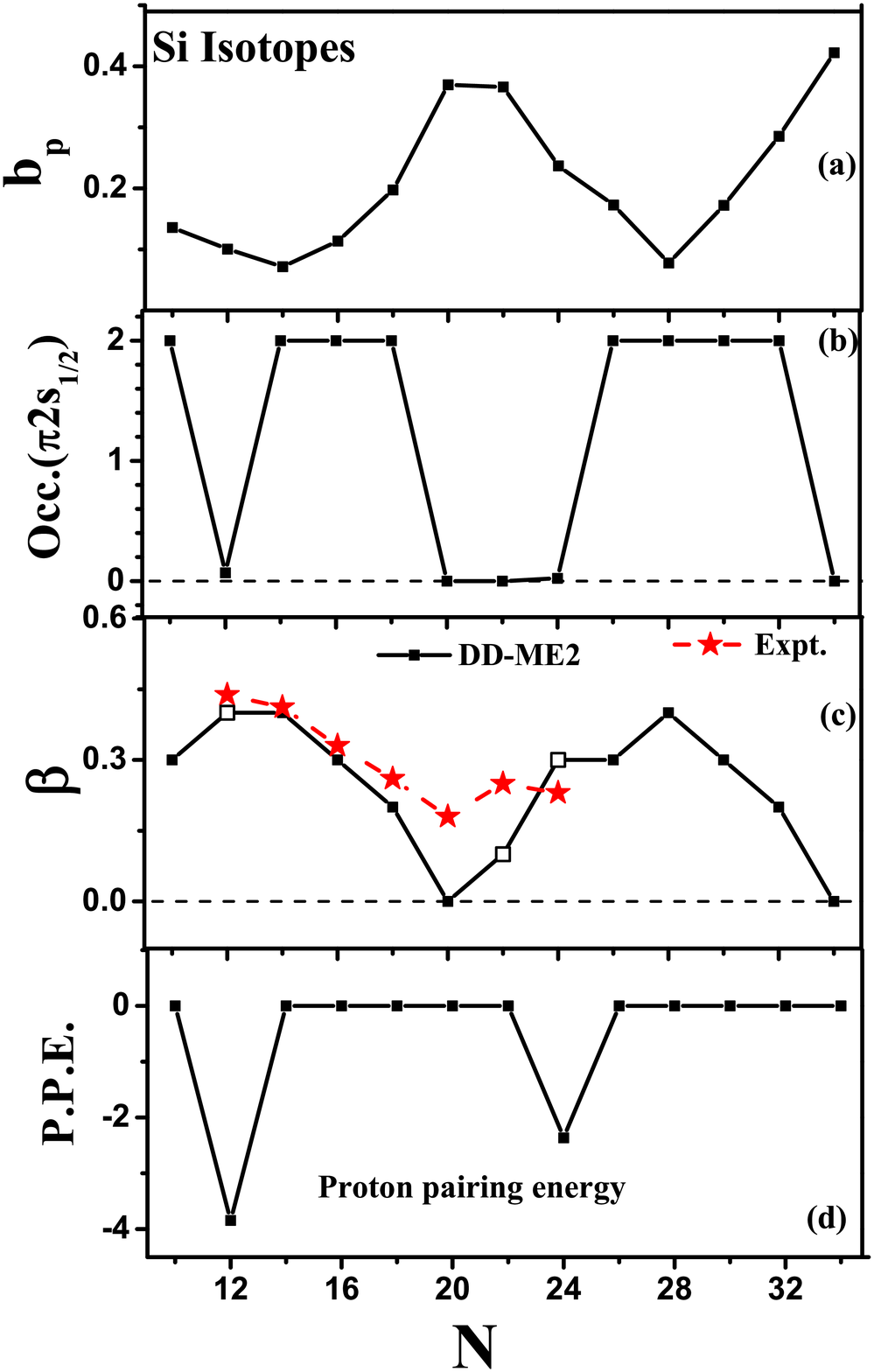}
\caption{Variations of bubble parameter $b_p$, occupancy of $2s_{1/2}$ state, deformation and pairing energy contribution (in MeV) as a function of
neutron number for full isotopic chain of Si.} \label{fig6} \end{figure}

There have been clear indications for a relationship
between the central density in nuclei and the symmetry energy
~\cite{Schuetrumpf17}. The central depression of proton density in
heavy nuclei is predominantly driven by Coulomb energy and the neutron
to proton ratio~\cite{Saxena18}. The pairing and dynamical correlations
associated with quadrupole shapes have been indicated to hinder the bubble
effect \cite{Khan07,Wu13,Wang11,Nakada12} in light nuclei. Moreover,
the reduced central density well below the nuclear matter saturation
value is expected to result in the loss of binding energy. This invokes
an investigation to explore the relationship between the central
depletion and various components of the binding energy viz. Coulomb
energy, pairing, deformation and isospin. It is important to point
out here that the role of Coulomb interaction is already demonstrated
in the superheavy bubble nuclei \cite{Saxena18}. Nuclear deformation
which changes with neutron to proton ratio and also influences the
central depletion (as shown in Figs. \ref{fig2} \&\ref{fig5}) is an
important parameter ~\cite{Duguet16,Khan07,Grasso09,Yao12,Wu13} to be
investigated systematically. Hence, we look into the sensitivity of the
bubble parameter to all these factors and study the full isotopic chain
of Si, which consists of a wide range of deformation with many oblate,
few prolate and spherical nuclei. Here, we plot and study the systematic
variation of (a) bubble parameter $b_p$ (b) occupancy (occ.) of $2s_{1/2}$
(c) quadrupole deformation ($\beta$) and (d) proton pairing energy
(P.P.E.) vs. neutron number $N = 10-34$ ($Z = 14, A = 24-48$) computed
using DD-ME2 parameter in Fig. \ref{fig6}.\par

In Fig. \ref{fig6}, the occupancy in $2s_{1/2}$ state is found zero for
spherical ($^{34,48}$Si) and prolate isotopes ($^{26,36,38}$Si, shown
by opaque symbol in Fig. \ref{fig6}(c)), whereas, it is fully occupied for oblate isotopes
similar to what found in Fig. \ref{fig5}. At magic neutron number $N =
20$, for the case of $^{34}$Si and at $N = 34$ for the case of $^{48}$Si,
which are found spherical ($\beta = 0$), the occupancy of $2s_{1/2}$ state
is zero and $b_p$ is maximum. However, the value of $b_p$ for $^{48}$Si
is even higher than the $b_p$ of the well known $N = 20$ ($^{34}$Si)
hinting a stronger bubble nature of $^{48}$Si as reported very recently
in Ref.$~$ \cite{Li2018}. A thorough inspection in various panels of
Fig. \ref{fig6} reveals that all the other parameters like proton pairing
energy (PPE $= 0$) contribution, deformation ($\beta = 0$) and occupancy
of $2s_{1/2}$ state ($= 0$) are same in both the nuclei $^{34}$Si and
$^{48}$Si. Therefore, the more $b_p$ for $^{48}$Si may be anticipated
due to neutron excess indicating 'neutron to proton ratio' is one factor
that also influences the central depletion and show higher $b_p$. This was
also seen in our recent work \cite{tbubble} in case of Ar isotopes where
proton rich $^{32}$Ar does not show central depletion, whereas $^{46}$Ar
has been found to exhibit bubble effect and also neutron rich $^{68}$Ar
has been indicated as bubble nucleus ~\cite{Grasso07,Khan07}.\par

On the other side, for neutron number $N = 28$ and also for $N = 14$,
where the deformation shows its maximum with oblate shapes, the $b_p$
is lowest which once again indicates the effect of oblate deformation
and full occupancy of $2s_{1/2}$, strengthening our outcome from
Fig. \ref{fig5}. The nuclear deformation shown in Fig. \ref{fig6}(c) for
the full chain of Si isotopes along with experimental data \cite{nndc}
show good agreement. While comparing the bubble parameter $b_p$
(Fig. \ref{fig6}(a)) and deformation $\beta$ values, we find that an
inverse trend between $b_p$ and $\beta$ authenticate the quenching effect
of deformation on bubble nuclei.  The proton pairing energy plotted in
Fig. \ref{fig6}(d) is found to be zero for all the Si isotopes except
for a few cases ($N = 12$ and $24$). Therefore, $Z = 14$ is found to
be a shell closure as per RMF approach for most of the isotopes and
hence gives doubly magic character to $^{34}$Si \cite{Mutschler16}
and $^{48}$Si \cite{Li16,Li2018}. Since the pairing correlations are
expected to quench the bubble, therefore, for $N = 12$ and $24$ the
non-zero pairing energy or rather the pairing correlations may be one
of the cause for the lower value of $b_p$ in nuclei $^{26,38}$Si as
reflected from Figs. \ref{fig6}(a) and (d). With a close watch of Fig.
\ref{fig6}, one can find that moving from $N = 10$ to $N = 12$, shape
changes from oblate to prolate and $2s_{1/2}$ state becomes fully occupied
to unoccupied which favour central depletion but the value of deformation
increases and also proton pairing energy (PPE) reaches non-zero which
consequently reduces the bubble parameter $b_p$. In a similar manner,
while moving from $N = 22$ to $N = 24$, even if shape remains prolate
and occupancy remains zero but value of deformation increases along with
non-zero contribution of PPE which lead to a sharp drop in the value
of $b_p$ from $^{36}$Si to $^{38}$Si. Hence the reduction of central
depletion is attributed due to the pairing and deformation correlations in
these cases. Therefore, this analysis shows that the central depletion is
actually a complex phenomenon, which is affected by the combined effects
of N to Z ratio, pairing correlations, deformation, and the occupancy
of $s_{1/2}$ state. Among these factors, the occupancy of $s_{1/2}$
state and the deformation influence significantly on the value of $b_p$.

\begin{figure}
\centering
\includegraphics[width=0.6\textwidth]{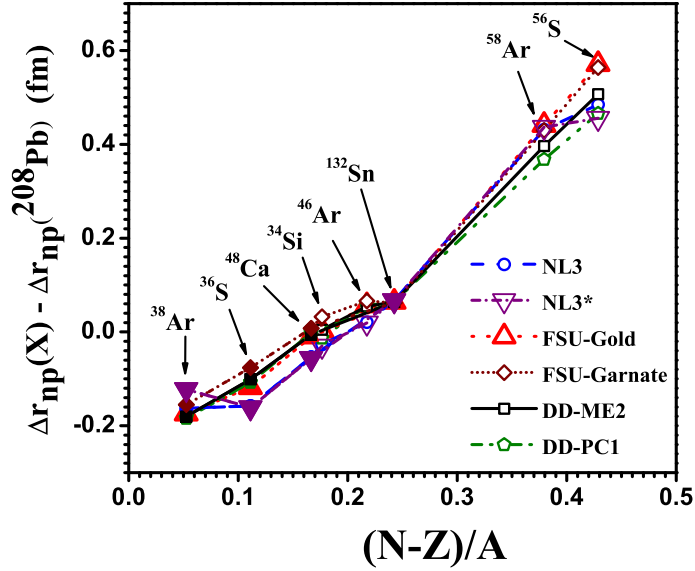}
\caption{(Colour online) $\Delta r_{np}(X)-\Delta r_{np}$($^{208}$Pb)
($\Delta r_{np}(X)$ is the neutron-skin thickness for nucleus $'X'$ in
question) plotted as a function of asymmetry $ [(N-Z)/A]$ obtained for
several RMF models as considered. The hollow symbols indicate nuclei
with central depletion in proton density and filled symbols are for
nuclei without central depletion.} \label{fig7} \end{figure}

Another important aspect relevant to the present study is to check the possible imprint of the central
depletion in the proton density on the systematics of the neutron-skin thickness, is presented here. The neutron-skin thickness $\Delta r_{np}$ is given by
\begin{equation}
\Delta r_{np} =  {\langle r_n^2\rangle}^{1/2} - {\langle r_p^2\rangle}^{1/2},
\end{equation}
which is the difference between the rms radii for the point neutrons and
the protons density distributions. Although the neutron skin thickness
is a surface effect, the central depletion in the proton density might
affect the rms radii for the proton distributions. Further, the change
in the proton density might also modify the neutron distributions in
the nuclei due to the self-consistency of the mean-field. The neutron
and proton densities at the center tend to be more or less the same
in the absence of the central depletion and the excess neutrons are
pushed to the surface causing the neutron-skin. However, one does not
know a priori, the influence of central depletion of proton density
on the neutron distributions or the neutron-skin thickness. One often
considers the variations of the neutron-skin thickness with the asymmetry
$(N-Z)/A$ parameter. The correlations among the neutron-skin thickness
of different nuclei are also usually considered. These systematics
enable one to assess the information content of the neutron-skin
thickness of a single nucleus. It is thus important to examine how
the systematics of the neutron-skin thickness might get influenced
in the presence of depletion of the central density in the nuclei.
In Fig. \ref{fig7}, we plot neutron-skin thickness as a function
of the asymmetry $(N-Z)/A$ parameter for several spherical nuclei
for different RMF models considered. The value of neutron-skin for a
nucleus is plotted in reference to its value for $^{208}$Pb nucleus for
a given model. It is interesting to note that the difference ($\Delta
r_{np}(X)-\Delta r_{np}$($^{208}$Pb)) linearly depends on the asymmetry
parameter irrespective of the model. The nuclei having central depletion
in the proton density (e.g., $^{34}$Si, $^{46,58}$Ar and $^{56}$S, see
Table \ref{tab2}) follow the trend similar to those having no central
depletion. Central density for the protons in the nuclei with central
depletion may be as small as half of its maximum density (i.e., b$_p$
$\sim$ 0.5).  The results presented in Fig. \ref{fig7} clearly suggests
that the depletion in the central density of the nuclei does not affect
the systematics of the neutron-skin thickness in a noticeable manner. \par

\begin{figure}
\centering
\includegraphics[width=0.6\textwidth]{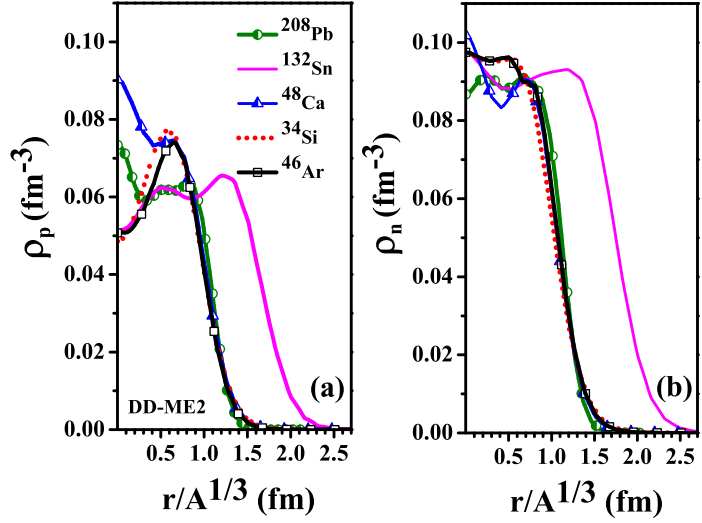}
\caption{(Colour online) Variation of proton and neutron densities
as a function of the radial co-ordinate $r$ scaled by $A^{1/3}$ for
the $^{34}$Si, $^{46}$Ar, $^{48}$Ca, $^{132}$Sn and $^{208}$Pb nuclei
obtained by DD-ME2.} \label{fig8} \end{figure}

For a better insight of the robustness of systematics of neutron-skin
thickness with respect to the central depletion, we compare the neutrons
and protons density profiles for $^{34}$Si and $^{46}$Ar with those of
nuclei $^{48}$Ca, $^{132}$Sn and $^{208}$Pb in Fig. \ref{fig8}. These
densities are plotted as a function of the radial co-ordinate $r$ scaled
by a factor of $A^{1/3}$. The density distributions for the protons
and neutrons for the $^{34}$Si and $^{46}$Ar show similar trends in
Fig. \ref{fig8}, where the central depletion in proton density for
$^{34}$Si and $^{46}$Ar is clearly seen which is at variance with the
densities of $^{48}$Ca and $^{208}$Pb. The effects of such depletion in
proton density are partially compensated by neutron density as can be
seen from Fig. \ref{fig8} (b). It may be noted that for the values of
$r \sim 0.75 A^{1/3}$ fm, neutron densities of $^{34}$Si and $^{46}$Ar
remain constant compared to neutron densities of $^{48}$Ca, $^{132}$Sn and
$^{208}$Pb nuclei. One might thus expect these differences in the density
distributions to affect the neutron-skin thickness in the bubble nuclei.

\begin{figure}
\centering
\includegraphics[width=0.6\textwidth]{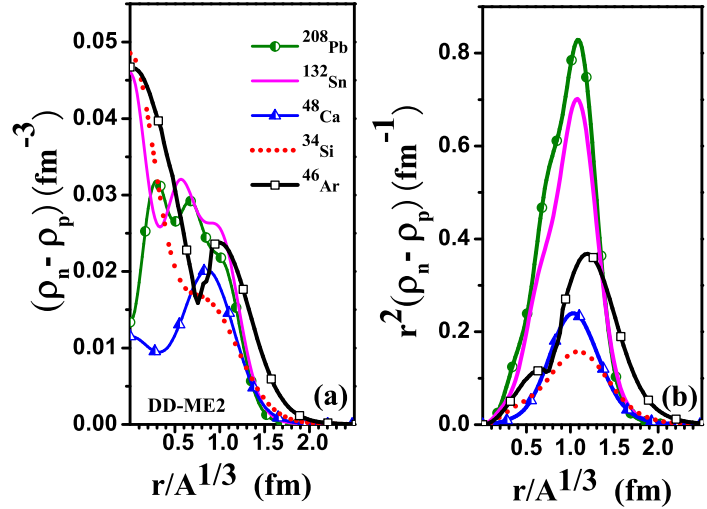}
\caption{(Colour online) The density difference $\rho_n$-$\rho_p$ and
r$^2$($\rho_n$-$\rho_p$) shown as a function of $r/A^{1/3}$ for $^{34}$Si,
$^{46}$Ar, $^{48}$Ca, $^{132}$Sn and $^{208}$Pb nuclei obtained for the
DD-ME2.} \label{fig9} \end{figure}

We plot the differences between the neutron and proton densities ($\rho_n
- \rho_p$) and $r^2(\rho_n - \rho_p)$ as a function of $r/A^{1/3}$ in
Fig. \ref{fig9}. The latter quantity may be more appropriate in order to
assess the influence of central depletion in the proton density on the
neutron-skin thickness $\Delta r_{np}$.  The differences between neutron
and proton densities $\rho_n - \rho_p$ for $^{34}$Si and $^{46}$Ar are
maximum at center, whereas it tends to be small in case of $^{48}$Ca and
$^{208}$Pb nuclei at the center. The behaviour of ($\rho_n -\rho_p$)
for $^{132}$Sn nucleus close to the center is similar to those for
$^{34}$Si and $^{46}$Ar nuclei, but at moderate value of $r/A^{1/3}$, it
follows the trend as those of $^{48}$Ca and $^{208}$Pb. The  differences
$r^2(\rho_n - \rho_p)$, however, look  pretty much the same for the
nuclei with or without the central depletion as can be noticed from
Fig. \ref{fig9}(b). The values of $r^2(\rho_n - \rho_p)$ peak around $r
\sim A^{1/3}$ fm. The peak heights are mainly governed by the number of
excess neutrons, i.e., $N - Z$. The dissimilarities in the $(\rho_n -
\rho_p)$, arising due to the central depletion, may not have any imprints
on the values of the neutron-skin thickness. In other words, the value
of the neutron-skin thickness is mainly governed by the differences
between the neutron and proton densities around the surface region and
the effects of the central depletion do not percolate that far.\par

\section{Conclusions}

The existence of depletion in the central density is explored in the
isotonic chains of nuclei with the neutron numbers $N = 20$ and $28$
using different variants of the relativistic mean-field model. The RMF
models considered are the ones (a) which include contributions from
the non-linear self- and mixed-interactions of the mesons with constant
coupling strengths and (b) with only linear contributions from the mesons
where the coupling strengths are density-dependent. The central depletion
of proton density is observed in several spherical as well as in deformed
nuclei and found to be affected not only by occupancy of $2s_{1/2}$ state
as usually believed, but, also by pairing correlations, deformation and
neutron to proton ratio. Bubble parameter shows the inverse dependence on
deformation and the occupancy in the s-orbit. The depletion in spherical
nuclei are found to disappear if the occupancy of $2s_{1/2}$ becomes more
than $\sim$ 50$\%$, while, it vanishes in deformed nuclei even though the
occupancy of $2s_{1/2}$ is almost zero. On the prolate side the bubble
parameter decreases even though the occupancy of $2s_{1/2}$ state is
almost zero whereas on oblate side, decrease in the bubble parameter is
associated with the combined effects of deformation and the increase in
occupancy of $2s_{1/2}$ state. \par

We find that the density-dependent point coupling model yields smaller central depletion in general. For instance, the density-dependent point coupling model results in practically no central depletion in the case of $^{46}$Ar nucleus, whereas other RMF models show strong central
depletion in the proton density. This model dependence is traced back to be associated with the differences in the strength of the spin-orbit potentials in these models.\par

The imprints of the central depletion on the systematics of the
neutron-skin thickness are investigated. It is found that the nuclei
with the proton central density as small as half of its maximum density
do not alter the systematics of neutron-skin thickness. The variations
of neutron-skin thickness with the asymmetry for the nuclei with central
depletion are very much in the harmony with those of the normal nuclei
(i.e., no central depletion). The  profiles for neutron and proton
densities for the nuclei with central depletion do seem to be at variance
with those of the normal nuclei. The effects of the central depletion,
however, do not percolate to the surface regions which contributes
maximally to the determination of the neutron-skin thickness.

\section*{ACKNOWLEDGMENTS}
G.S. and M.A. acknowledge the support by SERB for YSS/2015/000952 and
WOS-A schemes respectively.


\section*{References}

\begin{thebibliography}{99}
\bibitem{Mutschler16} A. Mutschler et al., Nature Phys. \textbf{13}, 152 (2017).
\bibitem{Todd04} B. G. Todd-Rutel, J. Piekarewicz, and P. D. Cottle, Phys. Rev. C \textbf{69}, 021301 (2004).
\bibitem{Grasso07} M. Grasso, Z. Y. Ma, E. Khan, J. Margueron, and N. Van Giai, Phys. Rev. C \textbf{76}, 044319 (2007).
\bibitem{Duguet16} T. Duguet, V. Som, S. Lecluse, C. Barbieri, and P. Navrtil, Phys. Rev. C \textbf{95}, 034319 (2017).
\bibitem{Schuetrumpf17} B. Schuetrumpf, W. Nazarewicz, and P.-G. Reinhard, Phys. Rev. C \textbf{96}, 024306 (2017).
\bibitem{Li16} J. J. Li, W. H. Long, J. L. Song, and Q. Zhao, Phys. Rev. C \textbf{93}, 054312 (2016).
\bibitem{Phuc2018} L. T. Phuc, N. Q. Hung, and N. D. Dang, Phys. Rev. C \textbf{97}, 024331 (2018).
\bibitem{Saxena17} G. Saxena, M. Kumawat, M. Kaushik, U.K. Singh, S.K Jain, S. Somorendro Singh, and Mamta Aggarwal, Int. J. Mod. Phys. E \textbf{26}, 1750072 (2017).
\bibitem{Saxena18} G. Saxena, M. Kumawat, M. Kaushik, S.K. Jain, and Mamta Aggarwal, Phys. Lett. B \textbf{788}, 1 (2019).
\bibitem{Khan07} E. Khan, M. Grasso, J. Margueron, and N. Van Giai, Nucl. Phys. A \textbf{800}, 37 (2008).
\bibitem{Sobiczewski2007} A. Sobiczewski and K. Pomorski, Progress in Particle and Nuclear Physics \textbf{58}, 292 (2007).
\bibitem{Decharge99} J. Decharg, J. F. Berger, K. Dietrich, and M. S. Weiss, Phys. Lett. B \textbf{451}, 275 (1999).
\bibitem{Singh12} S. K. Singh, M. Ikram, and S. K. Patra, Int. J. Mod. Phys. E \textbf{22}, 135001 (2012).
\bibitem{Ikram14} M. Ikram, S. K. Singh, A. A. Usmani, and S. K. Patra, Int. J. Mod. Phys. E \textbf{23}, 1450052 (2014).
\bibitem{Grasso09} M. Grasso, L. Gaudefroy, E. Khan, T. Niksic, J. Piekarewicz, O. Sorlin, N. V. Giai, and D. Vretenar, Phys. Rev. C \textbf{79}, 034318 (2009).
\bibitem{Yao12} J.-M. Yao, S. Baroni, M. Bender, and P.-H. Heenen, Phys. Rev. C \textbf{86}, 014310 (2012).
\bibitem{Wu13} X. Y. Wu, J. M. Yao, and Z. P. Li, Phys. Rev. C \textbf{89}, 017304 (2014).
\bibitem{Wang11} Y. Z. Wang, J. Z. Gu, X. Z. Zhang, and J. M. Dong, Chin. Phys. Lett. \textbf{28}, 10 (2011).
\bibitem{Nakada12} H. Nakada, K. Sugiura, and J. Margueron, Phys. Rev. C \textbf{87}, 067305 (2013).
\bibitem{Wang11a} Y. Z. Wang, J. Z. Gu, X. Z. Zhang, and J. M. Dong, Phys. Rev. C \textbf{84}, 044333 (2011).
\bibitem{Shukla2014} A. Shukla and S. Aberg, Phys. Rev. C \textbf{89}, 014329 (2014).
\bibitem{walecka74} J. D. Walecka, Ann. Phys. (N.Y.) \textbf{83}, 491 (1974).
\bibitem{Boguta77} J. Boguta and A. R. Bodmer, Nucl. Phys. A \textbf{292}, 413 (1977).
\bibitem{Boguta83} J. Boguta and H. Stoecker, Phys. Lett. B \textbf{120}, 289 (1983).
\bibitem{Furnstahl97} R. Furnstahl, B. D. Serot, and H.-B. Tang, Nucl. Phys. A \textbf{615}, 441 (1997).
\bibitem{Todd-Rutel05} B. G. Todd-Rutel and J. Piekarewicz, Phys. Rev. Lett \textbf{95}, 122501 (2005).
\bibitem{Lalazissis97} G. A. Lalazissis, J. Konig, and P. Ring, Phys. Rev. C \textbf{55}, 540 (1997).
\bibitem{Lalazissis09} G. Lalazissis, S. Karatzikos, R. Fossion, D. P. Arteaga, A. Afanasjev, and P. Ring, Phys. Lett. B \textbf{671}, 36 (2009).
\bibitem{Chen15} W.-C. Chen and J. Piekarewicz, Phys. Lett. B \textbf{748}, 284 (2015).
\bibitem{Lalazissis05} G. A. Lalazissis, T. Niksic, D. Vretenar, and P. Ring, Phys. Rev. C \textbf{71}, 024312 (2005).
\bibitem{Niksic08} T. Niksic, D. Vretenar, and P. Ring, Phys. Rev. C \textbf{78}, 034318 (2008).
\bibitem{Friedrich82} J. Friedrich and N. Voegler, Nucl. Phys. A \textbf{373}, 192 (1982).
\bibitem{nndc} http://www.nndc.bnl.gov/
\bibitem{Song2015} J. L. Song, Q. Zhao, and W. H. Long (2015), [arXiv:1504.04738] [nucl-th].
\bibitem{sharma95} M. M. Sharma et al., Phys. Rev. Lett. \textbf{74}, 3744 (1995).
\bibitem{satula} A. Bhagwat, R. Wyss, W. Satula, J. Meng and Y. K. Gambhir, arXiv: nucl-th/0605009v4.
\bibitem{Niksic14} T. Niksic, N. Paar, D. Vretenar, P. Ring, Comp. Phys. Comm. \textbf{185}, 1808 (2014).
\bibitem{moya} E. Moya de Guerra, P. Sarriguren, J. A. Caballero, M. Casas, D. W. L. Sprung, Nucl. Phys. A \textbf{529}, 68 (1991).
\bibitem{zamick} L. Zamick, D. C. Zheng, S. J. Lee, J. A. Caballero and E. Moya de Guerra, Ann. Phys. (N.Y.) \textbf{212}, 402 (1991).
\bibitem{Li2018}Jia Jie Li, Wen Hui Long, Jerome Margueron, Nguyen Van Giai, Phys. Lett. B \textbf{788}, 192 (2019).
\bibitem{tbubble} G. Saxena, M. Kumawat, B. K. Agarwal and Mamta Aggarwal, Phys. Lett. B \textbf{789}, 323 (2019).

\end{thebibliography}
\end{document}